# Advantages of rapid solidification over casting of Mg-0.4Zn-1Y alloy

Drahomír Dvorský[1*], Shin-Ichi Inoue[2], Ayami Yoshida[2], Jiří Kubásek[3], Jan Duchoň[1], Esther de Prado[1], Andrea Školáková[1], Klára Hosová[3], Petr Svora[1], Yoshihito Kawamura[2]

[1]Institute of Physics, Czech Academy of Science, Prague, Na Slovance 1999/2 182 21 Praha 8, Czech Republic

[2]Magnesium Research Center, Kumamoto University, Kumamoto, 2-39-1 Kurokami Chuo-ku Kumamoto, 860-8555 Japan

[3]Department of Metals and Corrosion Engineering, Faculty of Chemical Technology, University of Chemistry and Technology Prague, Technická 5 166 28 Praha 6 – Dejvice, Czech Republic

*dvorskyd@fzu.cz

**Abstract**. The Mg-Y-Zn magnesium alloy system is commonly recognized for its remarkable combination of high strength and ductility, achieved even with minimal amounts of alloying elements. This exceptional performance is attributed to its unique microstructure, which includes Long-Period Stacking Ordered (LPSO) phases or the distinctive microstructure derived from the LPSO phase, referred to as the Mille-Feuille structure (MFS). This study systematically compares the traditional ingot metallurgy method with the rapid solidification technique, coupled with diverse heat treatments and extrusion processes. Microscopic analyses reveal variations in the presence of LPSO phases, Mille-Feuille structure, and grain size, leading to divergent mechanical and corrosion properties. The rapid solidification approach stands out, ensuring superior mechanical properties alongside a reasonable corrosion rate.

## 1. Introduction

Magnesium materials are promising for biodegradable implants due to their favorable biocompatibility and mechanical properties, which closely resemble those of bone tissue [1]. However, enhancing the mechanical and corrosion properties of these materials is necessary to ensure their effective performance as supportive implants without adverse effects from corrosion products or released hydrogen, which could impede the healing process [2]. Additionally, utilizing magnesium alloys with minimal alloying elements is advantageous for sustaining biocompatibility in medical applications.

The demand for weight reduction is particularly pronounced in the automotive and aviation industries, where lower weight not only translates to decreased fuel consumption but also contributes to reduced $CO_2$ emissions. Among materials in consideration, magnesium alloys emerge as notable candidates due to their unparalleled lightweight characteristics coupled with acceptable mechanical properties [3]. Moreover, the greatest disadvantage of magnesium for aviation industry lies in the high affinity to oxygen as it may lead to disastrous consequences due to its low ignition temperature, high flame temperature and problematic extinguishing [4, 5]. Fortunately, some alloys exert increased ignition temperature and can pass flammability tests especially due to the presence of Y [6, 7]. That was the main reason behind the recently lifted ban by the Federal Aviation Administration (FAA) of magnesium alloys in the aircraft cabin [8]. Therefore, to effectively compete with materials such as aluminum or titanium, enhancements in mechanical, corrosion and ignition properties are necessary. Addressing this need, advancements can be achieved through the smart selection of alloys and a tailored preparation methodology. This proposition finds support in the work of Kawamura et al. [9], who successfully demonstrated the efficacy of this approach. Specifically, they employed rapid solidification to prepare the Mg-2Y-1Zn alloy, resulting in a high yield strength of 610 MPa combined with reasonable ductility, developing superior alloy to other high-strength magnesium and aluminum alloys at both room and elevated (200 °C) temperatures [10, 11]. This highlights the potential for optimizing the properties of magnesium alloys through precise alloy selection and innovative preparation techniques.

This pivotal discovery [9] started an in-depth exploration of the Mg-2Y-1Zn alloy and other magnesium alloys containing specific Long-Period Stacking Ordered (LPSO) phases. These phases exhibit a distinctive long-range periodic arrangement of atoms along the basal planes, typically spanning several nanometers, bestowing unique mechanical properties to the material. Extensive research has shown that LPSO phases contribute significantly to the enhancement of both strength and ductility [12-16]. The ability to form LPSO phases is possible after several conditions are met. Firstly, the reduction of stacking fault energy (SFE) is essential [17, 18]. Stacking faults are energetically favorable for solute segregation [19]. Y, when introduced, significantly reduces SFE, while Zn alone has a marginal effect on increasing SFE. Notably, Y and Zn together exert a synergistic effect, leading to a substantial reduction in SFE [20]. This lowered SFE enables the creation of wide GP zones by increasing the distance between leading and trailing Shockley partial dislocations [21]. The second crucial factor involves the formation of GP zones containing Y-Zn clusters [21-23]. Within the Mg-Y-Zn system, $L1_2$-type clusters, composed of $Zn_6Y_8$, are situated in-plane on the basal plane at the stacking faults, where they create four solute enriched atomic layers [22, 24-26] which are thermally stable due to the presence of Y or Mg atom in the center of the $L1_2$ cluster [27-30]. This synergy between solute element segregation and stacking faults is sometimes referred to as "synchronized LPSO structures" in order to distinguish them from simple LPSO structures [31]. The stability of LPSO is further influenced by the energetic disparity between the hexagonal close-packed (hcp) and face-centered cubic (fcc) structures. Alloying elements concentrate along the stacking faults, generating a local fcc lattice that coexists with the hcp structure due to the glide of the basal plane of the hcp structure with Shockley partial dislocation [17, 18, 32, 33]. First principle calculations support the notion that the transformation into an fcc structure is energetically favorable, as the energy gain associated with the transformation from $Y_6Zn_6 + Y_2$ clusters in an hcp lattice into $Y_8Zn_6$ $L1_2$ clusters sufficiently overcomes the energy loss caused by stacking-fault formation [34]. Kim et al. [21] utilized Transmission Electron Microscopy (TEM) to observe the precipitation of LPSO in Mg-Y-Zn alloy. They identified long plate-type defects along the basal plane with an increased concentration of Y and Zn. These were identified as GP zones with hcp structure, which transformed into fcc structure upon reaching a critical concentration of Y/Zn. Therefore, the separation distance between GP zones plays an important factor during the formation of the LPSO structure. Consequently, different treatments resulted in the observation of multiple types of LPSO phases, including 10H, 14H, 18R, and 24R, denoting rhombohedral or hexagonal lattice structures and atomic distances between each layer, with 14H being identified as the most stable configuration [21, 35, 36]. Except for Mg-Y-Zn alloying system LPSO phases were observed in another alloying systems containing combination of rare earth elements such as Sm, Gd, Tb, Dy, Ho, Er, or Tm with another metallic element such as Al, Co, Ni, or Cu [11, 37]. These alloying elements have their characteristic crystal structures, atomic radius, solubility in α-Mg, and mixing enthalpy between them [29, 31]. Moreover, there are two types of LPSO alloys, whether they crystallize from the melt (Type I) or whether they are not present in casted ingot and only precipitate from the α-Mg solid solution by heat treatment (Type II) [38].

Okuda et al. observed the formation of the LPSO phase from the amorphous alloy. Initially, α-Mg precipitated, followed by growth and arrangement of clusters to form LPSO phase [39, 40]. A single long plate-type defect, consisting of four layers with face-centered cubic (fcc) stacking (ABCA), wherein Y and Zn atoms are enriched in the successive B and C layers [19], has been designated as the LPSO building block by Kim et al. [21] and as the cluster arranged layer (CAL) by Kawamura et al. [41] as scanning tunneling microscopy revealed that clusters have a domain structure with a locally ordered arrangement [42]. Therefore, LPSO phase consists of hard CALs rich on alloying elements with constant distance (several nanometers) filled with soft Mg matrix between them. These phases are typically observed separately from magnesium grains, resulting in a material microstructure characterized by magnesium matrix and discrete LPSO phases, as documented by various researchers [43-46].

Even for Type I alloys it may be feasible to dissolve LPSO phases within the magnesium matrix through appropriate heat treatment followed by subsequent aging that facilitates their precipitation in irregular CALs within magnesium grains. Irregular CALs refer to variable distances between them, ranging from a single atomic layer to hundreds of nanometers. Kawamura et al. [41]

introduced the concept of cluster arranged nanoplates (CANaPs), comprising multiple CALs separated by up to four magnesium atomic layers. Consequently, post-heat treatment, the material is composed of magnesium grains containing multiple CANaPs with randomly varying distances between them. The spacing and thickness of CANaPs depend on the heat treatment conditions. The dispersion (quantity of CANaPs per unit distance) of CANaPs peaks after a specific aging duration, while the thickness gradually increases with aging time [41]. Studies have shown that the yield strength is significantly influenced by CANaPs dispersion, with no notable dependence on their thickness. Jian et al. [47] investigated the influence of stacking fault (SF) density and proposed a dependency similar to the Hall-Petch relation, expressed in the form of Equation (1).

$$\sigma_{0.2} = \sigma_a + k \cdot d^{-1} \qquad (1)$$

Where $\sigma_a$ represents the flow stress resulting from all strengthening mechanisms apart from parallel stacking faults, $k$ is a material constant, and $d$ is the mean spacing between stacking faults (SFs). Consequently, a higher density of SF structures contributes to increased strengthening [47]. Conversely, a denser quantity of precipitates can lead to reduced ductility [48].

Microstructures featuring CANaPs have been previously observed by various authors [17, 21, 35, 49], they were often identified as LPSO phases due to the shared basic building block. In recent literature, these structures are denoted as Mille-Feuille structures (MFS) owing to the repetitive layers of soft magnesium matrix and hard CALs [41, 50, 51]. The MFS microstructure offers a unique advantage by combining the hardness of LPSO phases with the ductility of the matrix [52].

Mechanical properties after heat treatment alone remain suboptimal, therefore thermomechanical processing is necessary. Strength enhancement in magnesium alloys is achieved through solid solution strengthening, the Hall-Petch relation, dispersion strengthening, and texture strengthening [19]. Extrusion, the most common processing method for magnesium, is closely tied to dynamic recrystallization (DRX) enhanced by the particle-stimulated nucleation mechanism [53]. Consequently, a fine-grained material is achieved, and strength is improved in accordance with the Hall-Petch relation.

Depending on the extrusion conditions, worked regions with prolonged, and deformed non-recrystallized grains (non-DRX) are typically present. In magnesium materials, there is a tendency for basal planes to align parallel with the extrusion direction, resulting in a strong basal texture in non-DRX grains [54]. This texture is usually associated with significant asymmetry in compressive yield strength (CYS) and tensile yield strength (TYS) due to the lower energy requirement for twinning than the slip mechanism as compression readily induces twinning while tension suppresses it [46, 55]. Although DRX grains may also exhibit basal texture, twinning becomes more challenging with reduced grain size [56]. LPSO and MF structures impede non-basal slip dislocations [19]. The MF structure employs a unique deformation mechanism called kinking, which suppresses twinning and results in low asymmetry of mechanical properties [19]. Additionally, during extrusion, the MF structure introduces kinks into the material, leading to kink strengthening [41, 46]. A higher dispersion of kinks correlates with increased strength [41]. While a strong basal texture tends to reduce ductility, in the case of the MF structure, ductility is partially preserved. Otherwise, the high ductility of this alloy is typically associated with small grains exhibiting random orientation.

The behavior of a microstructure containing LPSO phases, MF structure, or a rich solid solution during extrusion is distinct, leading to varied final properties dependent on both the initial microstructure and extrusion conditions. Numerous publications highlight the advantageous properties resulting from rapid solidification or the utilization of metal powders for compaction, as opposed to traditional casted ingots [9, 57, 58]. Rapid solidification methods prove beneficial due to the swift cooling process, resulting in supersaturated solid solutions and the formation of very fine grains. However, challenges may arise concerning particle consolidation, such as the presence of oxides on the surface or potential alterations to the microstructure due to elevated temperatures during consolidation. Nevertheless, these inhomogeneities might even present new opportunities for material preparation methods [59].

This paper focuses on the comparison of conventional casting with a novel rapid solidification method. Different heat treatments yield various microstructures containing LPSO phases or MF structure, which significantly influence the final properties after extrusion.

## 2. Materials and methods
### 2.1. Material preparation

Alloy with composition of Mg-0.38Zn-1Y (at. %) was prepared by melting of pure elements in induction furnace at 750 °C under protective argon atmosphere. The melt was casted into steel mold with 30 mm in diameter. Two ingots with final composition of Mg-0.39Zn-0.99Y (at. % according to the ICP measurement) were prepared. One ingot was separated into 3 pieces, where one was labeled as I (As cast ingot), the second and third piece underwent the solid solution treatment at 540 °C for 72 h under protective argon atmosphere followed by quenching in water. The second piece was labeled as IT4 (solid solution treatment). Finally, the third piece was aged at 350 °C for 12 h and it was labeled as IT6 (aged). The second ingot was re-melted and rapid solidification was performed under protective argon atmosphere. For rapid solidification the melt at temperature of 800 °C was injected on the cooled brass disk with rotations of 2000 RPM. Rapidly solidified (RS) ribbons were compressed in the copper billets with the pressure of 25 MPa. Degassed copper billets filled with compressed RS ribbons were then hot pressed at 350 °C at 100 MPa for 10 min followed by cooling in water. Copper surface was removed from the bulk material. Such billet was labeled as RSHP (rapid solidification, hot press) Solid solution heat treatment was performed in the furnace under protective Ar atmosphere at temperature of 540 °C for 72 h followed by quenching in water. This state is referred to as RSHP-T4. Aging treatment was performed at 350 °C for 12 h for all samples. This state is referred to as RSHP-T6. The ingots and green compact samples were then extruded at the same machine at 375 °C with extrusion ratio 7.5 and ram speed of 2 mm·s$^{-1}$. Summary of sample processing and labeling prior and after extrusion (same extrusion conditions for all samples) is given in **Table 1**.

**Table 1:** Sample processing and labeling prior and after extrusion.

| Before extrusion | Solidification/processing | Heat treatment | After extrusion |
|---|---|---|---|
| I | Casting | - | IEx |
| IT4 | Casting | 540 °C/72h | IT4Ex |
| IT6 | Casting | 540 °C/72h + 350 °C/12h | IT6Ex |
| RS | Rapid solidification | - | - |
| RSHP | Rapid solidification + hot press 350 °C | - | RSEx |
| RSHPT4 | Rapid solidification + hot press 350 °C | 540 °C/72h | RST4Ex |
| RSHPT6 | Rapid solidification + hot press 350 °C | 540 °C/72h + 350 °C/12h | RST6Ex |

### 2.2. Microstructure characterization

For microstructure characterization the samples were ground on the SiC papers P80-P4000, afterward, they were polished on diamond pastes 3 μm and 1 μm. The final polishing was performed on the Eposil F suspension. Samples were etched in the solution of 4.2 g picric acid, 10 ml H$_2$O, 10 ml acetic acid, 70 ml ethanol in order to investigate the grain size. The grain size was determined by image analysis of 3 images in imageJ. Microstructure was studied by confocally optical microscopy (OM; OPTILICS HYBRID). SEM measurements with EDS analysis was done using a TescanVEGA3 LMU scanning electron microscope (SEM) equipped with energy dispersive spectroscopy (EDS). The electron back-scattered diffraction (EBSD) analysis was performed using an FEI 3D Quanta 3D field-emission-gun DualBeam scanning electron microscope equipped with an EBSD detector TSL/EDAX Hikari. EBSD mapping was performed with a step size ranging from 30nm to 0.1 μm, acceleration voltage of 20 kV, and beam current of 32 nA. Data were analyzed using a EDAX OIM Analysis 8 software. An observation of the microstructure was per formed by Conventional Transmission Electron Microscopy (CTEM) and Scanning Transmission Electron Microscopy (STEM) with a High Angle Annular Dark Field (HAADF) detector using a FEI Tecnai G2 F20 X- TWIN microscope with a double-tilt specimen holder operated at 200 kV. X-ray diffraction (XRD) analysis was performed using PANalytical X'Pert Pro in Bragg-Brentano geometry with a Co anode.

### 2.3. Mechanical properties

Compressive and tensile properties were measured using INSTRON 1362 machine at room temperature at a strain rate of 0.001 s$^{-1}$. Cylindrical samples with a diameter of 5 mm and 7.5 mm high were used for compressive tests. Compressive yield strength (CYS) and ultimate compressive strength (UCS) were determined from compressive curves. Tensile tests were performed on dog bone specimens with a gauge length of 25 mm and diameter equal to 3.5 mm. Tensile yield strength (TYS), ultimate tensile strength (UTS), and elongation to fracture (A) were determined from compressive curves. Three measurements were performed for each material.

*2.4. Corrosion properties*

Corrosion behaviour was studied in the simulated body fluid (SBF) prepared according to Müller [60]. The starting pH was set to 7.4 at 37 °C. An SBF volume of 100 ml per 1 $cm^2$ of sample surface was used. Samples were fixed in plastic holders and immersed in SBF for 14 days at 37 °C. The corrosion rate was evaluated based on the weight changes after the removal of corrosion products. Three measurements were performed for all materials.

*2.5. Ignition temperature*

Samples with dimensions 5x5x5 mm were used for ignition tests. Samples were put into the $Al_2O_3$ crucible which was inserted into the chamber of a resistance furnace. One thermocouple was in the direct contact with the sample and the other one was situated in the middle of the crucible. The tests were performed at a stable flow of technical air (100 $l·h^{-1}$). The heat rate corresponded to the 25 °C·$min^{-1}$. The temperature regularly increased with time during the test until there was rapid increase due to the ignition of the sample.

## 3. Results

*3.1. The microstructures*

Casting is a prevalent method for preparing various materials, including magnesium. **Figure 1A** illustrates the microstructure of a casted ingot of Mg-0.38Zn-1Y alloy, revealing large grains ranging from 100 to 1500 µm and a characteristic dendritic microstructure. The dendrites consist of magnesium solid solution and a eutectic intermetallic phase. Micro-segregation within the magnesium solid solution is also evident. X-ray diffraction (XRD) analysis (**Figure 2**) confirms that this material comprises α-Mg solid solution and a eutectic ternary Long-Period Stacking Ordered (LPSO) phase. The patterns have been converted to the Cu wavelength for better comparison with the literature.

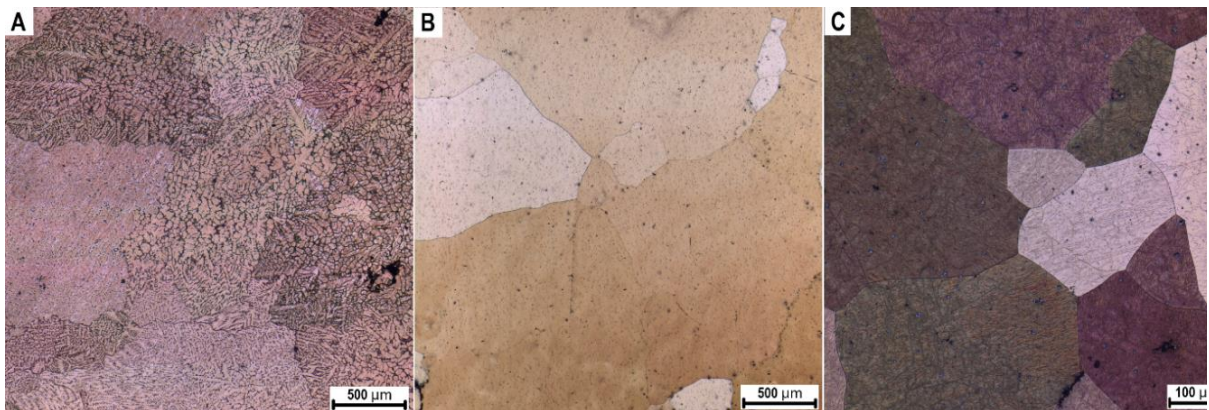

**Figure 1**: Microstructures of A) casted ingot (I), B) ingot after solid solution treatment (IT4), C) ingot after aging (IT6)

After solid solution treatment of the casted ingot, the eutectic phase dissolves into the magnesium solid solution (**Figure 1B**). However, a few thermally stable phases, particularly $YH_2$, emerged in the microstructure. The grain size remained similar, with sizes ranging between 100 and 1500 µm. Subsequent aging of the ingot maintains these thermally stable phases while introducing a specific arrangement of Y and Zn-rich layers (**Figure 1C**). Notably, the grain size remains almost unchanged during the aging process.

The ribbons produced through rapid solidification exhibit a thickness of approximately 30 µm and a width of about 800 µm (**Figure 3A**). Notably, the side closer to the brass wheel features smaller grains (0.1 – 2 µm) attributed to the rapid solidification process. Conversely, the opposite side displays slightly elongated larger grains (2 – 10 µm). Rapid solidification, owing to its swift cooling, prevents the formation of secondary phases, and all alloying elements remain dissolved in the solid solution.

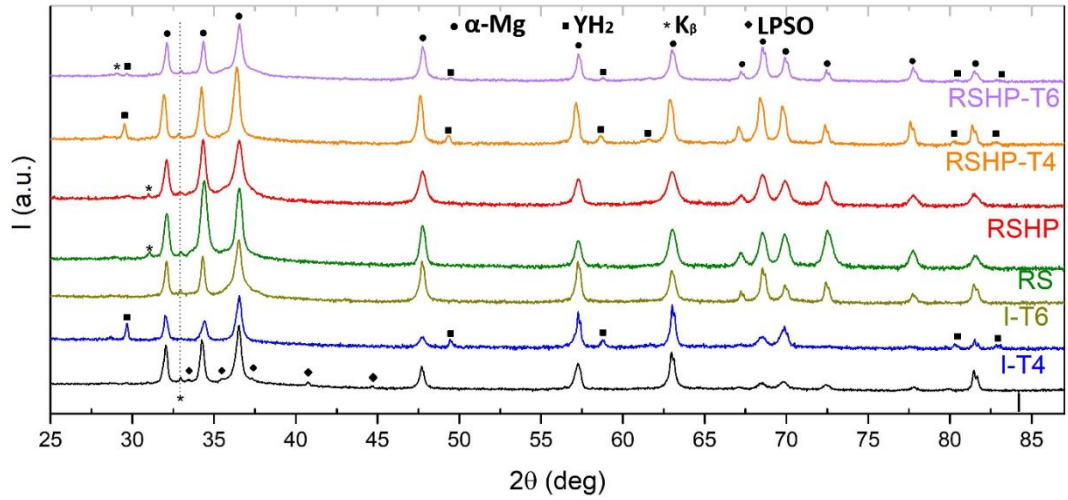

**Figure 2**: XRD analysis of samples before extrusion.

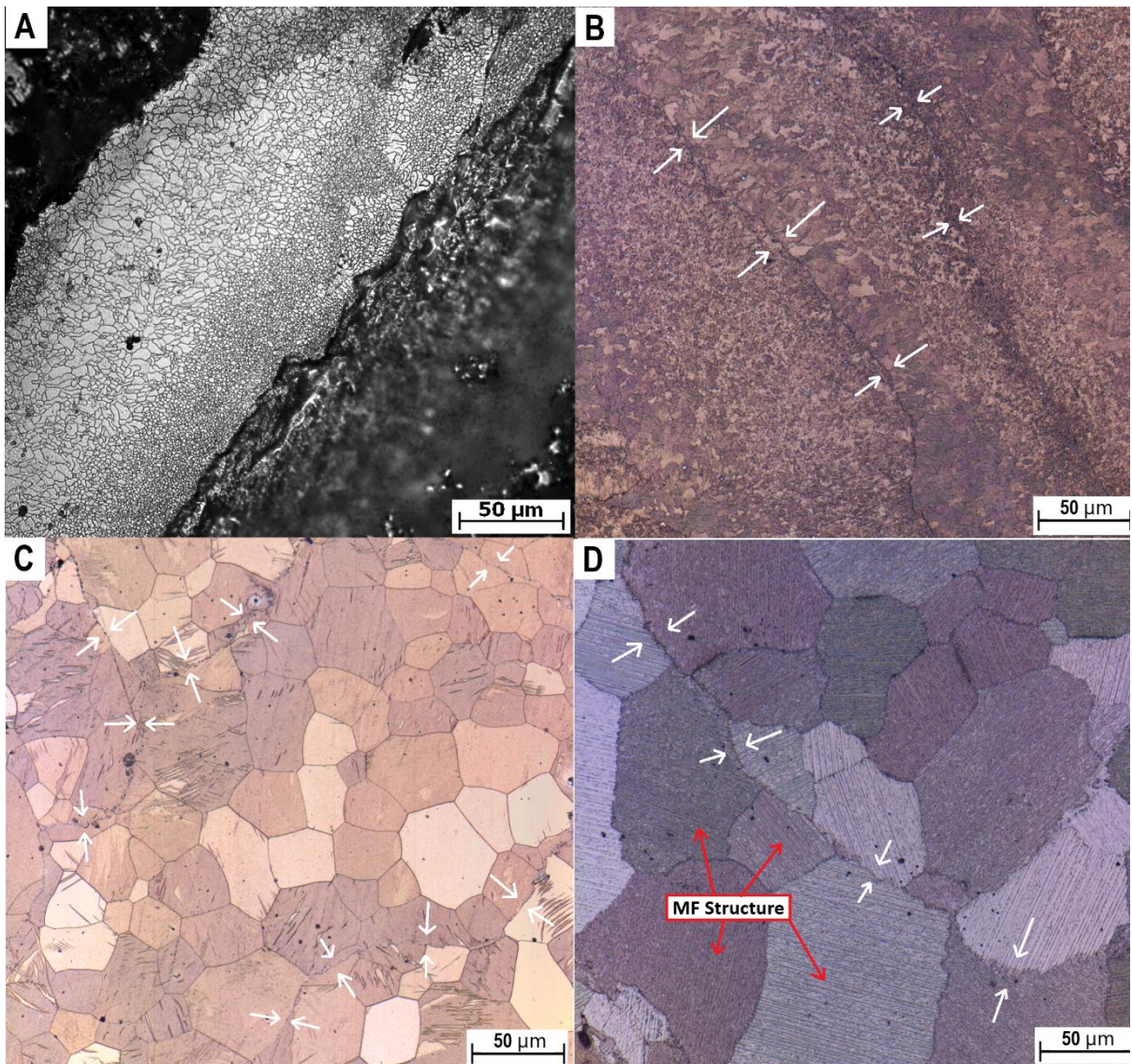

**Figure 3**: Microstructures of A) rapidly solidified ribbon (RS), B) hot pressed billet (RSHP), C) billet after solid solution treatment (RSHP-T4), D) billet after aging (RSHP-T6) (white arrows indicating oxide layer)

The hot-pressed billet underwent a solution treatment, resulting in significant grain growth, as shown in **Figure 3C**. The absence of obstacles for grain growth, such as intermetallic phases or precipitates, facilitated easy grain growth. The only potential obstacle could be the thin oxide shell (indicated by white arrows). The range of grain size after solid solution treatment was approximately 5 – 100 μm. Importantly, all alloying elements remained dissolved in the solid solution. The microstructure occasionally featured $YH_2$ phases, which formed during the heat treatment.

The rapidly solidified ribbons underwent a series of processing steps: handled in a glove box, cold pressed, sealed, degassed, and finally hot pressed. The resulting microstructure is depicted in **Figure 3B**. Notably, the grain size after hot pressing remained similar to that of the rapidly solidified ribbons, with only negligible observed grain growth. The difference in grain size between the two sides of the ribbon is still evident. Additionally, a thin oxide shell is noticeable between each ribbon after hot pressing (indicated by white arrows). However, all alloying elements remain dissolved in the solid solution due to the relatively low temperature of pressing and the short exposure time.

Subsequent aging of the billet led to additional slight grain growth (10 – 110 μm) due to the elevated temperature, as illustrated in **Figure 3D**. However, owing to the supersaturated solution resulting from rapid solidification and solid solution treatment, Y and Zn precipitated in a specific manner. Each grain now contained nanoplates with a uniform orientation. These nanoplates look like scratches after etching, however, those "scratches" have different orientation in each grain and represent microstructure commonly known as Mille-Feuille (MF) due to the repeating layers of soft Mg matrix and hard Y and Zn-rich layers.

The casted ingot exhibits relative stability at the extrusion temperature of 375 °C, with no significant phase transformations occurring. The microstructure after extrusion is intricate, comprising elongated non-DRX grains (80 – 1100 μm) with strong basal texture, dynamically recrystallized equiaxed grains (0.5 – 8 μm) with weak basal texture (**Figure 5A**), and elongated intermetallic phases that facilitate the recrystallization process (**Figure 4A**). Consequently, approximately 44% of the material comprises dynamically recrystallized (DRX) areas (**Table 2**).

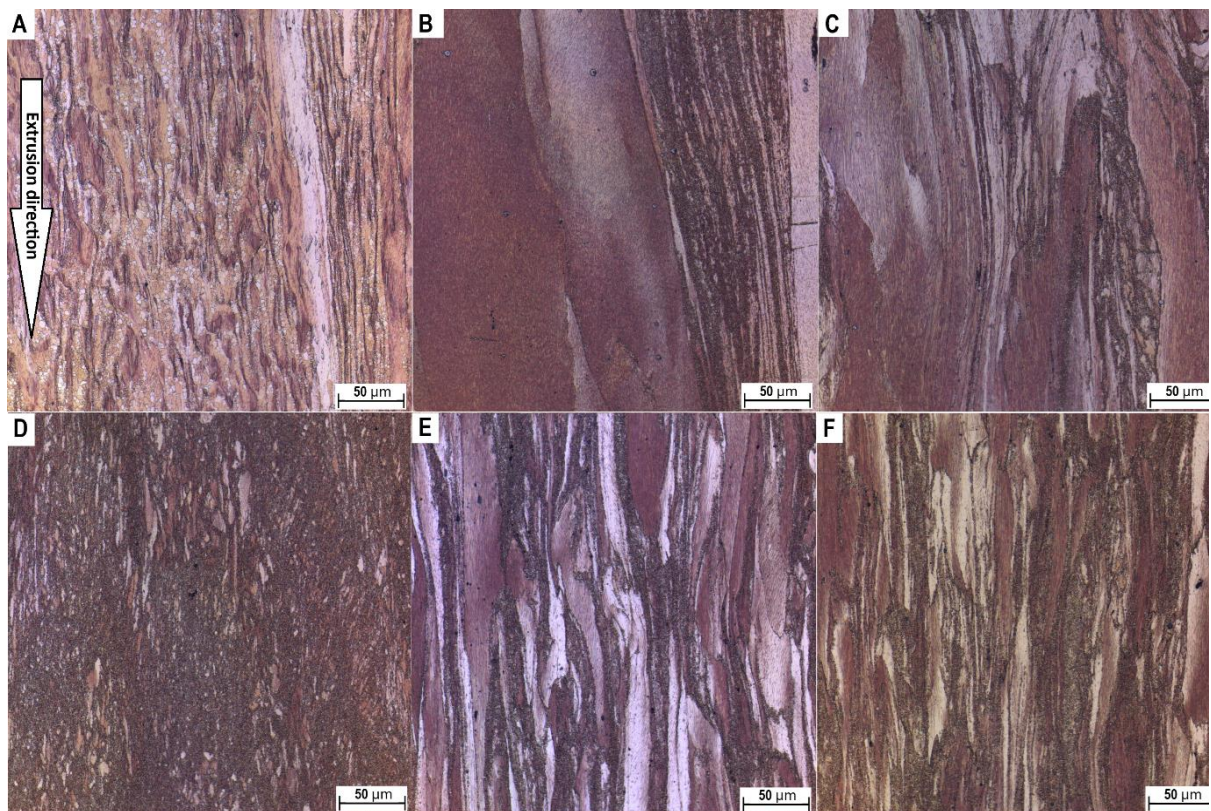

**Figure 4**: Microstructures (longitudal) after extrusion of A) casted ingot (IEx), B) solution treated ingot (IT4Ex), C) aged ingot (IT6ex), D) RSHP billet (RSEx), E) RSHP billet after solution treatment (RST4Ex), F) RSHP billet after aging (RST6Ex)

In contrast, the ingot after solid solution treatment is metastable, leading to precipitation during extrusion (**Figure 4B**). Even a brief exposure to elevated temperatures during extrusion is sufficient to enable precipitation. Consequently, extruded ingots after heat treatments contain MF structure and are characterized by very large, elongated grains (80 – 800 µm) with strong basal texture and significantly smaller stripes of recrystallized grains (0.5 – 3 µm) with weak basal texture (**Figure 5B**). There is also a notable decrease in the dynamically recrystallized (DRX) area compared to the IEx sample (**Table 2**).

**Table 2:** Microstructure attributes of extruded samples. DRX = recrystalized area [%], ED = equal diameter [µm].

|  | Ingot metallurgy | | | Rapid Solidification | | |
| --- | --- | --- | --- | --- | --- | --- |
|  | IEx | IT4Ex | IT6Ex | RSEx | RST4Ex | RST6Ex |
| DRX | 44 ± 5 | 30 ± 4 | 38 ± 4 | 85 ± 2* | 33 ± 5 | 37 ± 3 |
| ED of non-DRX | 459 ± 271 | 466 ± 271 | 445 ± 203 | 8 ± 3 | 66 ± 31 | 67 ± 29 |

*Fraction of finer grains

The microstructure of rapidly solidified billets after hot pressing is also metastable, with alloying elements in solid solution. During extrusion, some of these alloying elements precipitate into MFS. Consequently, this sample contains areas of very fine recrystallized grains (0.1 – 1 µm) and areas of larger elongated grains (2 – 12 µm), occasionally containing MFS (**Figure 4D**). This sample is characterized by the overall smallest grain size and the highest fraction of DRX grains (**Table 2**). Distinguishing between fine DRX and fine non-DRX grains is challenging; therefore, the given number represent a fraction of finer grains.

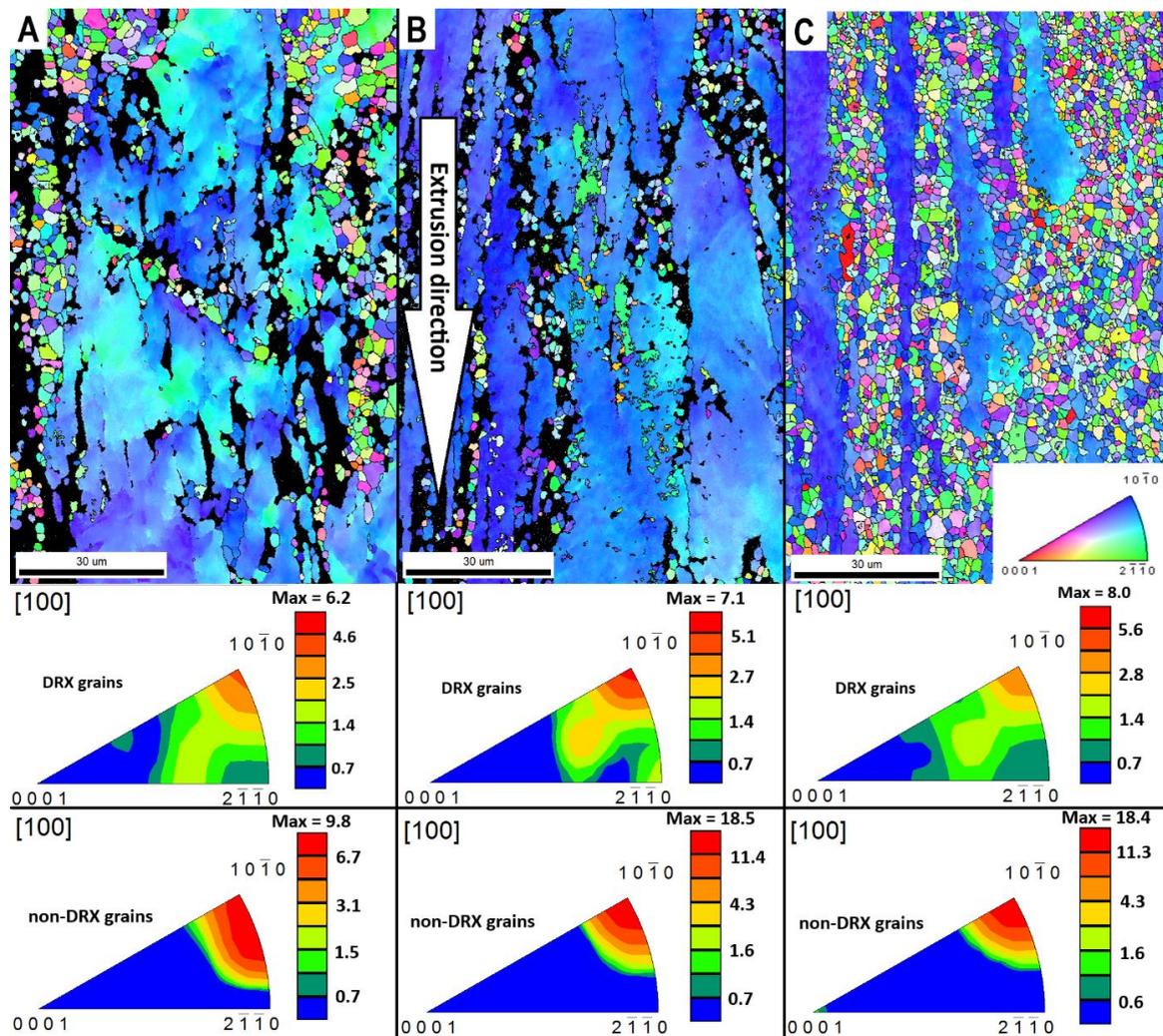

**Figure 5**: EBSD and inverse pole figures of DRX and non-DRX grains of A) IEx B) IT6Ex C) RST6Ex (black spots are unassigned points or LPSO phase)

The sample after solid solution treatment is metastable before extrusion, and due to the elevated temperature, some phases precipitate into MFS (**Figure 4E**). Dynamically recrystallized grains (0.5 – 3 μm) with weak basal texture are located along the elongated grains (15 – 250 μm) with strong basal texture. There is a lower content of DRX grains, but it is comparable with the heat-treated ingot, although the distribution differs significantly due to smaller non-DRX grains (**Table 2**).

Only the sample after aging is thermally stable during extrusion, as it already contains fully precipitated MFS (**Figure 4F**). Dynamic recrystallization also occurs along the elongated non-DRX grains with MFS and strong basal texture, and the grain sizes are very similar to the previous specimen (15 – 250 μm), although there is a slightly higher content of DRX grains with weak basal texture (**Figure 5C** and **Table 2**).

STEM observation of RST6Ex sample (**Figure 6A**) confirmed the presence of non-DRX grains with high density of dislocations and the presence of fine DRX region. Fine DRX grains (**Figure 6B**) also contain MF structure, however, their lines are strait in contrast with the bend lines of non-DRX grain (**Figure 6C**) due to deformation during extrusion.

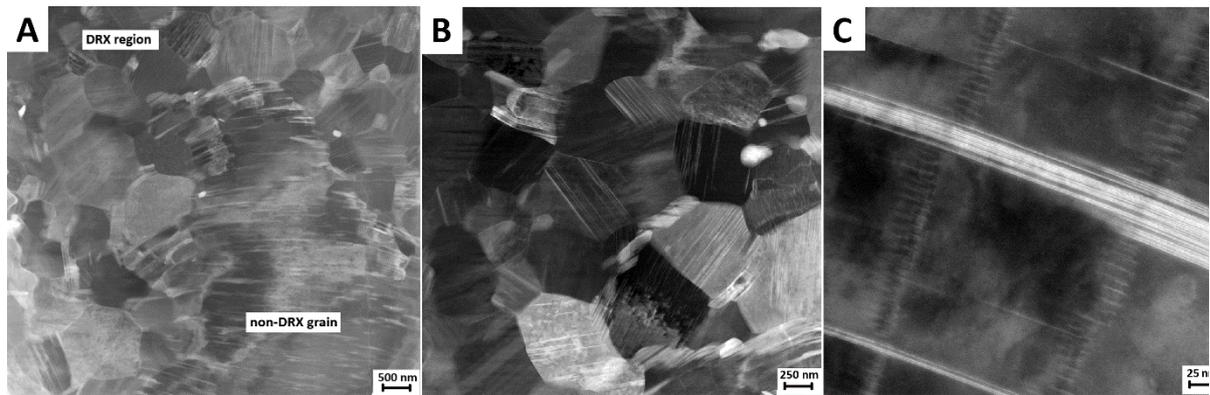

**Figure 6:** STEM images of cross section of RST6Ex (extrusion direction is perpendicular to the image area): A) overall image, B) detail of DRX region, C) detail of non-DRX grain.

### 3.2. Mechanical properties

Mechanical properties in tension and compression parallel to the extrusion direction were measured, and the values of compressive yield strength (CYS), ultimate compressive strength (UCS), tensile yield strength (TYS), ultimate tensile strength (UTS), and ductility (A) were evaluated, as summarized in **Table 3**. The ingot metallurgy product from the casted ingot exhibits a reasonable TYS of 290 ± 8 MPa and an elongation of 7.7 ± 1.6%. However, there is a significant difference of 43 MPa between CYS and TYS. Solid solution treatment leads to a significant improvement in both TYS and CYS by 30 and 28 MPa, respectively. However, it also results in a notable reduction in elongation by about 6%. Similar results are observed after aging. Nevertheless, there is a much lower difference between CYS and TYS (28 MPa), and a slightly higher elongation of 2.7 ± 0.8%.

**Table 3:** Mechanical properties of extruded samples.

|  | Ingot metallurgy | | | Rapid Solidification | | |
|---|---|---|---|---|---|---|
|  | IEx | IT4Ex | IT6Ex | RSEx | RST4Ex | RST6Ex |
| CYS [MPa] | 247 ± 3 | 275 ± 2 | 292 ± 5 | 388 ± 15 | 313 ± 19 | 334 ± 3 |
| UCS [MPa] | 412 ± 23 | 408 ± 16 | 533 ± 16 | 415 ± 1 | 533 ± 7 | 547 ± 16 |
| TYS [MPa] | 290 ± 8 | 320 ± 8 | 320 ± 7 | 367 ± 1 | 359 ± 4 | 353 ± 5 |
| UTS [MPa] | 322 ± 9 | 329 ± 8 | 334 ± 12 | 373 ± 3 | 371 ± 4 | 363 ± 2 |
| A [%] | 7.7 ± 1.6 | 1.8 ± 0.3 | 2.7 ± 0.8 | 8.0 ± 1.1 | 12.3 ± 1.5 | 12.8 ± 1.2 |

Products made by rapid solidification are notably superior to ingot metallurgy, both in tension and compression. The CYS can reach up to 292 MPa in the case of ingot metallurgy, while it reaches up to 388 MPa with rapid solidification. Similarly, TYS goes up to 320 MPa in ingot metallurgy compared to 367 MPa for rapid solidification. Contrarily, elongation is above 8% for all rapidly solidified products, while it is below 8% for ingot metallurgy. Otherwise, direct extrusion of hot-pressed billets leads to the highest strength, albeit with the lowest elongation among rapidly solidified products.

Mechanical properties of heat-treated products were similar, with only minor differences in CYS and elongation.

*3.3. Corrosion properties*

Corrosion tests were conducted in simulated body fluid (SBF) at a temperature of 37 °C in closed containers over a duration of 14 days. The corrosion rate was calculated based on the weight changes after exposure and removal of corrosion products according to equation (2), and the values are summarized in **Table 4**.

$$v_{cor} = \frac{365 \Delta m}{14 S \rho} \qquad (2)$$

Where $\Delta m$ is weight change, S is the samples surface area, $\rho$ is density of magnesium, and numbers refers to the days of exposition. All samples exhibited corrosion rates between 0.6 and 1 mm·a$^{-1}$, which falls within the optimal range for applications in medicine as biomaterials. The highest corrosion rate was measured for the directly extruded casted ingot, while the corrosion rate of heat-treated ingots was lower. On the other hand, the heat-treated RS products reached higher corrosion rate compared to directly extruded product which was investigated as the most resilient.

**Table 4:** Corrosion rates of prepared materials in SBF.

|  | Ingot metallurgy | | | Rapid Solidification | | |
|---|---|---|---|---|---|---|
|  | IEx | IT4Ex | IT6Ex | RSEx | RST4Ex | RST6Ex |
| $v_{cor}$ [mm·a$^{-1}$] | 0.99 ± 0.01 | 0.73 ± 0.03 | 0.68 ± 0.04 | 0.66 ± 0.02 | 0.84 ± 0.02 | 0.84 ± 0.01 |

*3.4. Ignition properties*

Ignition properties of extruded samples were measured in furnace with the presence of technical air and are summarized in **Figure 7**. The results show that ignition temperatures ranged between 880 and 950 °C showing only slightly higher ignition temperature of ingot metallurgy products compared to rapid solidification. Where the highest ignition temperature was measured for IT4Ex sample.

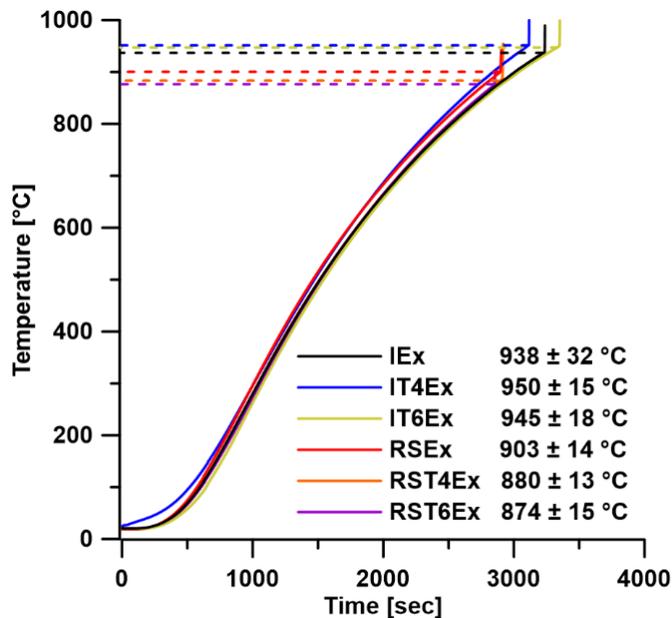

**Figure 7**: Measurements of ignition temperature of extruded samples.

## 4. Discussion

*4.1. Microstructure*

The microstructure of the casted ingot exhibited a typical coarse-grained dendritic microstructure with LPSO phases (**Figure 8A**), consistent with observations and descriptions by other researchers [43-46]. The solid solution treatment, followed by quenching in water, effectively dissolved all phases. Only thermally stable YH$_2$ phases were created during this heat treatment, a common occurrence in the thermal treatment of magnesium alloys containing Y [61]. The quenching process resulted in an increased amount of alloying elements in the solid solution [62].

Subsequent aging led to the specific precipitation of phases in rows (**Figure 8B**), a phenomenon associated with the low stacking fault energy of this alloy due to the presence of Y and Zn [17, 18, 20, 43]. The initial formation of GP zones with hcp structure along the basal planes was followed by their transformation into local fcc structures after reaching a certain Y/Zn concentration [21]. Consequently, Y and Zn formed Cluster Arranged Nanoplates (CANaPs) with uneven distances between them, known as the Mille-Feuille (MF) structure [22, 24, 41]. The detail of one CANaP after deformation can be seen in **Figure 6C**, where some of the individual CALs were being created but not finished.

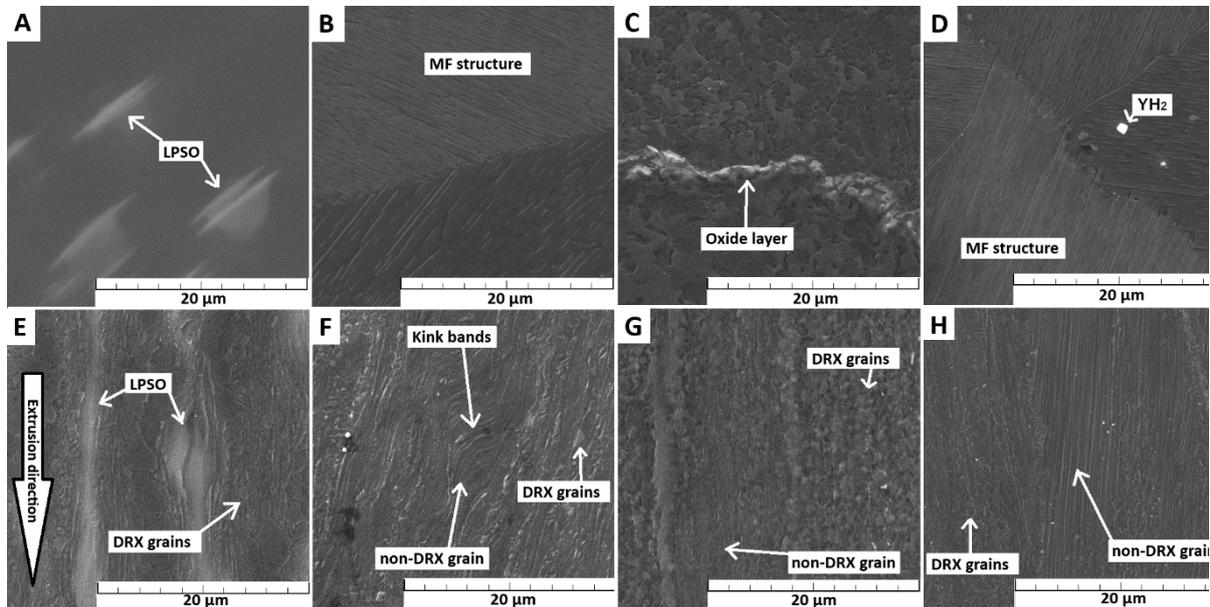

**Figure 8**: SEM images of details of specific microstructure features (top line – samples before extrusion, bottom line – samples after extrusion): A) I, B) IT6, C) RSHP, D) RSHP-T6, E) IEx, F) IT6Ex, G) RSEx, H) RST6Ex

Subsequent extrusion of the casted ingot (**Figure 8E**) resulted in a microstructure comprising magnesium solid solution, LPSO phases, small, recrystallized grains with weak texture, prolonged non-DRX grains with preferential basal orientation, and local MF structures with very low CANaP density. These features have been consistently observed in previous studies [44, 46, 55, 62]. The magnesium solid solution was not homogeneous, and microsegregations from the casted ingot persisted. LPSO phases fragmented and aligned along the extrusion direction, forming kink bands [45]. The presence of LPSO phases facilitated the recrystallization process, resulting in the highest volume of DRX regions among ingot products. Hagihara et al. [62], who prepared a similar alloy by extrusion, also observed improved recrystallization along the LPSO stripes, which were ordered in rows parallel to the extrusion direction. Easier recrystallization along LPSO phases is associated with the localization of stress at the interface between the Mg matrix [62]. Xu et al. also observed preferential recrystallization near LPSO phases [19]. However, large, non-DRX grains with a uniform diameter size of up to 1100 μm remained in the microstructure due to the relatively low extrusion temperature and ratio. The size of non-DRX grains is associated with the initial microstructure grain size. These non-DRX grains exhibited a typical basal texture, a well-known characteristic of magnesium materials after extrusion [46]. They contained a less prominent MF structure with occasional kinks. Local MF structure is formed by the segregation of Y and Zn in the stacking faults and at the grain boundaries, which occur along the basal planes during the elevated temperature of extrusion, similarly to aging [44]. In contrast, DRXed grains exhibited only a weak texture due to particle-stimulated nucleation [19].

The microstructure after heat treatments appears completely different compared to the extruded casted ingot. However, it is similar after solid solution treatment and aging. In both cases, the microstructure consists of non-DRX grains and DRX stripes of very fine grains, predominantly created at the original grain boundaries, as there was no LPSO phase to aid in the recrystallization process [62]. Additionally, some fine grains are observed at kink grain boundaries, as noted by Wu et al. [19]. The large prolonged non-DRX grains are associated with large initial grains, and these prolonged grains contain MF structure with kinks (**Figure 8F**) in both cases, even though there were no MF phases before extrusion. This phenomenon might be associated with the elevated temperature during extrusion, which is similar to the temperature of aging. However, due to the limited time at this temperature, the precipitation of the MF structure was not entirely completed; therefore, the IT4 sample contained fewer kinks and DRX regions, as MF structure also promote recrystallization [19].

On the other hand, rapid solidification yields very fine-grained ribbons, with one side of the ribbon consisting of finer grains, while the other side consists of slightly larger grains due to the different solidification rates close to the brass wheel and on the opposite side. Moreover, this material consists solely of a rich magnesium solid solution. Further processing of ribbons by hot pressing joined these ribbons together, where each ribbon was surrounded by a thin oxide layer (**Figure 8C**). This oxide layer served as a barrier for grain growth during the subsequent heat treatment. Otherwise, all alloying elements remained dissolved in the solid solution. Subsequent aging resulted in the specific precipitation of Y and Zn predominantly at the stacking faults, forming what is known as the MF structure (**Figure 8D**). A detailed investigation of the microstructure of rapidly solidified ribbons after heat treatment was conducted by Matsuda et al. [49], ensuring the preparation of a heat-stable and uniform material for extrusion.

Subsequent extrusion of the RSHP sample resulted in a microstructure with very fine DRX grains, as well as fine non-DRX grains (**Figure 8G**). This outcome is a consequence of the initial microstructure before extrusion, where very fine grains deformed and remained fine even after extrusion. Nevertheless, all non-DRX grains were surrounded by finer DRX grains, as recrystallization occurred predominantly at the original grain boundaries. DRX regions were created predominantly at the grain boundaries of the original grains, as well as some kink grain boundaries, which was also observed by Wu et al. [19]. Dynamic recrystallization is associated with particle-stimulated nucleation, as nanoplates end at grain boundaries [19]. The MF structure partially appeared in the larger non-DRX grains due to the metastable microstructure of the RS sample and elevated temperature during extrusion. However, it was difficult to distinguish between DRX grains and fine non-DRX grains; therefore, in this case, the DRX region was considered as grains less than 2 μm.

The microstructure of the extruded solid solution-treated sample differed from the RS sample. However, it was again similar to the extruded sample after aging. These materials were characterized by large, non-DRX grains surrounded by very fine DRX regions. In the case of solid solution treatment, there was likely not a complete precipitation of the MF structure due to the short time at elevated temperature during extrusion. This resulted in a slightly lower amount of DRX regions, as it seems that recrystallization occurs predominantly at the grain boundaries, which might be associated with the presence of nanoplates of the MF structure that end at the grain boundaries [19]. These spikes might act as nuclei for grain recrystallization. This behavior is beneficial, as each non-DRX grain should be surrounded by a DRX region (**Figure 8 H**), resulting in the presence of a microstructure resembling the harmonic microstructure [63-65].

*4.2. Mechanical properties*

For the discussion of mechanical properties, several values were calculated and are presented in **Table 5**. The CYS/TYS ratio describes the asymmetry of mechanical properties in compression and tension. A value closer to 1 indicates minimal asymmetry of mechanical properties. The UCS/CYS and UTS/TYS ratios describe the relative difference between yielding and maximum strength, corresponding to the potential for hardening. This can be further quantified by the hardening capacity ($H_{CT}$) and hardening capacity compression ($H_{CC}$) for tension and compression, respectively [66-68]. They are calculated using the equation (3).

$$H_C = \frac{US - YS}{YS} \qquad (3)$$

Where *US* is ultimate strength (UCS for compression (Hcc) or UTS for tension (Hct)) and *YS* is yield strength (CYS for compression (Hcc) or TYS for tension (Hct)). Higher hardening capacity means better ability of the material to harden during plastic deformation. Similar output can be delivered from the work hardening exponent *n* calculated by equation (4).

$$\sigma_T = K\varepsilon_T^n \qquad (4)$$

Where $K$ is strength coefficient, $\sigma T$ is true stress and $\varepsilon T$ is true strain. Therefore, the slope of the logarithmic curve of true stress versus true strain is used to determine the work hardening coefficient $n$.

The mechanical properties in compression and tension are influenced by the microstructure. Relative to the amount of alloying elements, the mechanical properties of most materials are very satisfactory for both aviation and medicine. The extruded casted ingot exhibited the lowest strength but also reasonable ductility. Generally, the strength of materials is associated with grain size, solid solution, and intermetallic phases. In this case, strength is improved by the presence of relatively fine and homogeneously dispersed LPSO phases. Conversely, if the LPSO phases are large, it may lead to a deterioration of mechanical properties, as observed by Liu et al. [45] who measured lower TYS of 272 MPa for extruded Mg-2Zn-4Y alloy with poor elongation of 2.8%. However, the main strengthening factor in this case comes from the non-DRX grains due to texture strengthening.

**Table 5:** Mechanical properties of extruded samples.

|  | Ingot metallurgy | | | Rapid Solidification | | |
|---|---|---|---|---|---|---|
|  | IEx | IT4Ex | IT6Ex | RSEx | RST4Ex | RST6Ex |
| UCS/CYS | 1.67 | 1.48 | 1.83 | 1.07 | 1.70 | 1.64 |
| UTS/TYS | 1.11 | 1.03 | 1.04 | 1.02 | 1.03 | 1.03 |
| CYS/TYS | 0.85 | 0.86 | 0.91 | 1.06 | 0.87 | 0.95 |
| $H_{cT}$ | 0.110 | 0.028 | 0.044 | 0.016 | 0.033 | 0.03 |
| $H_{cc}$ | 0.668 | 0.484 | 0.825 | 0.094 | 0.703 | 0.638 |
| $n$ | 0.080 | 0.018 | 0.042 | 0.032 | 0.052 | 0.089 |

After extrusion, the LPSO phase and worked Mg grains orient their basal planes (0 0 0 1) parallel to the extrusion direction. Therefore, the Schmid factor for the most common basal slip in these grains is small, limiting basal slip significantly. This means that LPSO phases and Mg worked grains are strong reinforcing components in tension [46, 62]. This specific orientation has consequences on the measured TYS, which tends to be high. Generally, in most materials, the CYS is usually higher compared to the TYS as during compression, some material defects (voids, cracks, etc.) are neglected and have a lower impact compared to tension. However, in most magnesium materials, CYS is usually lower compared to TYS due to the presence of basal texture after deformation. This orientation is favorable for twinning if the compressive force is applied [46, 55]. Twinning requires less energy than the slip of dislocations, resulting in lower CYS. The twinning mechanism also strongly depends on the grain size [69]. Twinning is hindered in magnesium materials with grain sizes lower than several microns [56]. In this case, the texture of IEx sample resulted in a significant difference of 43 MPa between CYS and TYS or CYS/TYS ratio in **Table 5**, which is the highest in the studied materials. The extensive twinning of large non-DRX grains can be observed on the microstructure after compression (**Figure 9A**). However, the texture strengthening is usually associated with the loss of ductility. Nevertheless, in this case, the elongation remained high, associated with the presence of a large volume fraction of DRX grains. These grains were characterized by a weak texture, improving ductility [62]. Moreover, these DRX grains were slightly larger compared to other samples, resulting in the highest UTS/TYS ratio, indicating some hardening during testing, more observable in the hardening capacity in tension ($H_{CT}$) and a high hardening exponent ($n$).

The measured CYS of the ingot metallurgy product (247 ± 3 MPa) was slightly lower compared to that measured by Hagihara et al. [62], who reported a CYS of 270 MPa for extruded Mg-1Zn-2Y alloy. This difference is associated with the lower amount of alloying elements and DRX regions in our case. According to the Hall-Petch relation, finer grains generally contribute to higher strength. However, in this case, the main contribution to TYS comes from the non-DRX grains due to texture strengthening, which, as discussed earlier, results in a lower CYS due to the twinning.

Samples IT4Ex and IT6Ex exhibited very similar microstructures, with the main difference being the amount of MF structure in non-DRX grains. Consequently, their mechanical properties were also similar. These materials displayed higher strength compared to the extruded ingot but with much lower ductility, making them less suitable for biomaterial applications. The low ductility observed can be attributed to the large, non-DRX grains and a lower content of DRX regions. The non-DRX grains were large and sometimes adjacent to other non-DRX grains, creating large areas of non-DRX grains and only stripes of DRX regions. Although the total DRX area was relatively close to that of the IEx sample with good elongation, the distribution of DRX and non-DRX regions played a more crucial role in total elongation. Ductility reduction is also influenced by texture strengthening, which was the main contributor to the strength increase compared to the extruded ingot due to the higher content of non-DRX grains. The strength improvement in IT4Ex and IT6Ex samples is influenced by solid solution strengthening, as solute Y atoms induce higher solid solution strengthening than other alloying elements [70]. Therefore, a higher content of Y in the solid solution leads to higher strength [62]. The mechanical properties of IT4 and IT6 samples exhibited slight differences. IT6 showed higher CYS and higher elongation, attributed to the higher content of nanoplates. The higher density of the MF structure leads to higher strengthening, surpassing the strengthening observed in materials containing LPSO phases [19, 47]. However, denser MF structure should also result in lower ductility [48]. The ductility was regained due to higher DRX content. While extensive twinning in non-DRX grains reduces the CYS of the IT4Ex sample (**Figure 9B**), the presence of MF structure in IT6Ex sample suppresses the formation of twins, and kinking becomes the dominant deformation mechanism as can be seen in microstructure after compression in **Figure 9C**. [46, 49]. This phenomenon requires more energy, compensating for the lower CYS [46, 55]. Additionally, the introduction of kinks, for example during extrusion, significantly increases the strength of the material [41, 46]. As a result, the CYS/TYS ratio is closer to 1 in the IT6Ex sample with a fully developed MF structure, effectively restricting twinning [19, 35]. There is also a higher hardening capacity compared to the IT4Ex sample.

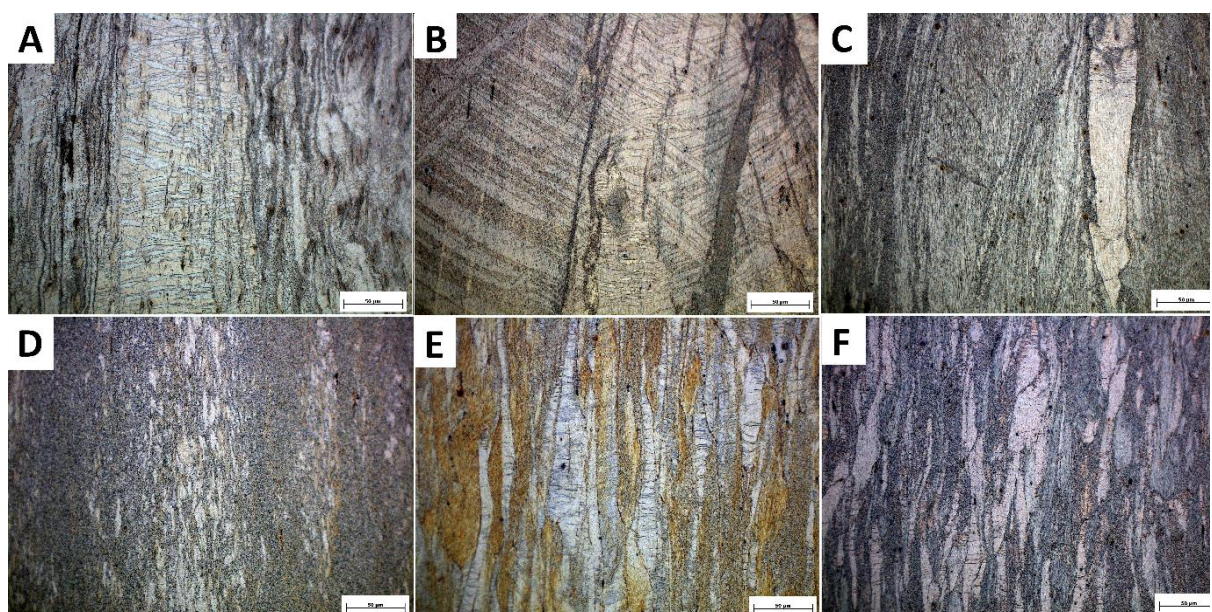

**Figure 9**: OM images of microstructure after compressive test: A) IEx, B) IT4Ex, C) IT6Ex, D) RSEx, E) RST4Ex, F) RST6Ex.

Samples prepared by rapid solidification demonstrated significantly higher strength and ductility compared to those obtained through traditional ingot metallurgy methods. Direct extrusion of the RSEx sample yielded the highest strength due to the smallest grain size and a texture of non-DRX grains containing MF structure with kinks. The presence of non-DRX grains (up to 15 µm) surrounded by a mixture of fine non-DRX grains (up to 2 µm) and fine DRX grains (up to 2 µm) created a harmonic microstructure, resulting in high ductility. Moreover, the RSEx sample was the only material characterized by a CYS/ TYS ratio higher than 1, indicating the ultimate suppression of twinning due to the presence of the MF structure and small grain size (**Figure 9D**) [49, 56, 69]. However, not all grains finer than 2 µm were likely DRXed, leading to a very low hardening capacity.

RST4Ex and RST6Ex samples had a similar microstructure, with a comparable correlation as observed in the case of heat-treated samples made from the ingot. The lower amount of nanoplates and the texture of non-DRX grains had a greater impact on the measured mechanical properties of the RST4Ex sample, resulting in lower CYS and elongation due to easier twinning (**Figure 9E**). This observation is reflected in the CYS/TYS ratio, as well as the hardening capacity and exponent. Out of the obtained results it seems that rapid solidification leads to the superior mechanical properties compared to ingot metallurgy, making this technique interesting for industry.

*4.3. Corrosion properties*

The measured corrosion rates for all tested samples are considered ideal for biomaterials, as an ideal corrosion rate is typically around 1 mm·a$^{-1}$ [71-73]. The low corrosion rate of this alloy is attributed to the presence of yttrium, which forms a stable $Y_2O_3$ layer on the sample's surface [74-76]. This layer can be more easily formed from the solid solution or less stable phases [77]. The concentration of yttrium in the corrosion products was observed in all samples, with more intense concentration in the heat-treated products (**Figure 10**).

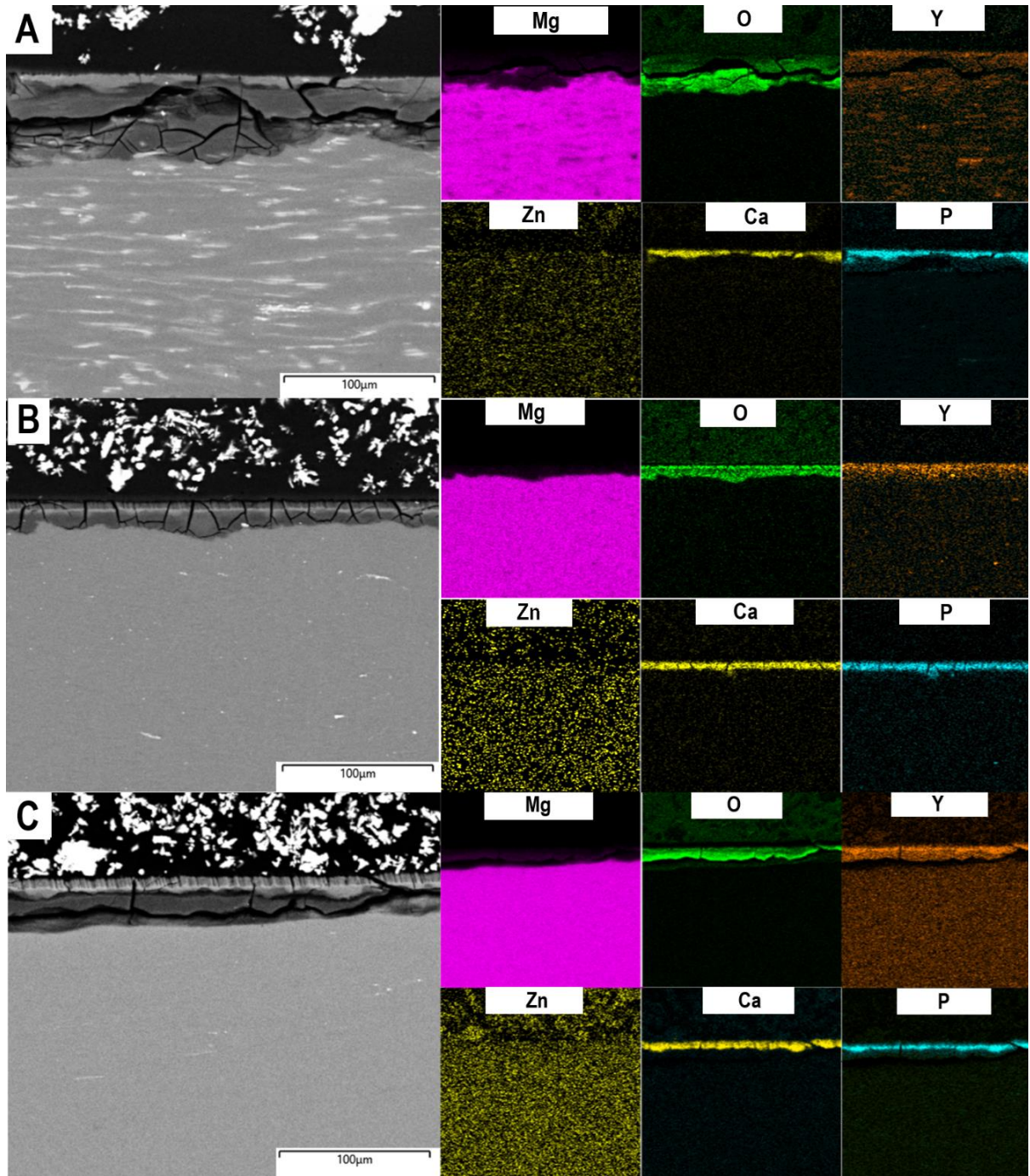

**Figure 10**: Cut of sample after corrosion with EDS analysis of corrosion products: A) IEx, B) IT6Ex C) RST6Ex

In the corrosion products there were also additions of Ca and P that precipitated from the SBF in the form of dicalcium phosphate or its dehydrate as Ca/P ratio is close to 1 [78]. The corrosion layer can be therefore divided into outer layer with majority of Ca and P elements, while the inner layer contained $Y_2O_3$ and $Mg(OH)_2$. The precipitation of Ca and P elements is useful in the application as biomaterials as it enhances the fixation of the implant to the bone and subsequent replacement by bone tissue [78].

The differences in the corrosion rate in individual samples can be described as follows. The highest corrosion rate of IEx sample is associated with the presence of secondary LPSO phase, partial microsegregation, and an overall inhomogeneous microstructure. Generally, fine grains lead to better corrosion resistance unless there is some microsegregation at the grain boundaries [79-81]. DRX grains contributed to relatively low corrosion rate. On the other hand, the presence of LPSO phases had much higher impact in material containing higher fraction of them due to higher content of alloying elements [16, 82]. By the dissolution of LPSO phase and segregated areas during the heat treatment the decrease of corrosion rate can be achieved as the content of Y in the solid solution increases [77]. The sample subjected to aging displayed a high density of MFS and increased DRX content, contributing to a further reduction in the corrosion rate. Conversely, the RSEx sample exhibited the lowest corrosion rate, attributed to its very fine microstructure. Subsequent heat treatment, leading to larger grain sizes, resulted in an increased corrosion rate. Despite having smaller non-DRX grain sizes than ingot metallurgy products, the corrosion rate in RSEx was slightly higher. This observation may be linked to the bimodal microstructure, where non-DRX grains contained MF structure, while DRXed regions were MF-free. This created slight potential differences, enhancing corrosion in non-DRX grains. In the case of ingot metallurgy, the larger local area of less resistant non-DRX grains compared to small DRX regions suppressed the microscopic effect of a small anode and large cathode. Conversely, RS products tended to have smaller non-DRX grains, bringing them into closer proximity to DRX regions. This allowed for a small anode effect, resulting in a slightly higher corrosion rate [83].

*4.4. Ignition properties*

The key factors influencing ignition resistance primarily include the type and quantity of alloying elements. Certain rare earth elements such as Nd, Sm, Yb, Dy, Y, Gd, and among non-rare earth elements Ca, and Sr are known to have a beneficial effect due to their ability to protect the melt surface with stable oxides [6, 84, 85]. Y alone can provide sufficient protection to magnesium, owing to its lower Standard Gibbs free energy for oxide formation compared to magnesium, which facilitates the replacement of the non-protective MgO layer with a $Y_2O_3$ layer [6]. Ignition temperatures can reach up to 818 °C with a Y content of 3 wt.% [5]. Rapidly solidified materials with a Y content of 3.5 wt.% achieved ignition temperatures between 874 and 903 °C. However, other factors such as Y content in the solid solution may also influence ignition resistance. $Y_2O_3$ formation is more favorable on the surface from the solid solution rather than intermetallic phases, similar to corrosion behavior. Interestingly, casted ingot materials exhibited slightly higher ignition temperatures. This could be attributed to the processing of the material, as rapidly solidified materials undergo multiple heat treatments during which oxidation occurs, depleting Y to form $Y_2O_3$. This $Y_2O_3$ is implemented in the material where it acts as a barrier against grain growth, however, reducing the Y available for surface oxide formation. In contrast, ingot metallurgy products do not contain $Y_2O_3$ within the material, leading to slightly higher ignition temperatures. Casted ingot exhibited higher deviation in values due to low homogeneity. Nevertheless, both casted ingots and rapidly solidified materials achieve acceptable ignition temperatures comparable to WE43 alloy [86, 87]. Moreover, the ignition properties of this alloy might be significantly improved by only a small addition of Be (0.1 at.%) [88].

## 5. Conclusion

The magnesium alloy Mg-0.4Zn-1Y was processed through conventional casting and rapid solidification, with subsequent heat treatments (T4 and T6) revealing the formation of the MF structure in both preparation methods. Before extrusion, notable differences existed in grain size, impacting the microstructure after extrusion. Ingot products exhibited large prolonged non-DRX grains surrounded by very fine recrystallized grains. The RS products shared a similar microstructure but with significantly smaller non-DRX grain sizes, resulting in a more homogeneous distribution of DRX regions.

Mechanical properties of the RS products surpassed those of ingot metallurgy, primarily due to smaller grain size and better DRX region distribution. DRX regions significantly contributed to the materials' ductility, thanks to the randomized texture of newly formed grains. Strength was associated with texture strengthening in non-DRX grains containing regular or irregular MF structures with kinks. Products lacking regular MF structure exhibited minor asymmetry in mechanical properties due to the basal texture of the non-DRX grains. Materials subjected to aging with regular MF structure suppressed twin formation in compression, leading to a CYS/TYS ratio closer to 1.

All prepared materials exhibited corrosion rates suitable for bioapplications. The presence of LPSO phases in the casted ingot correlated with the highest measured corrosion rate, while a homogeneous microstructure with very fine grains resulted in the lowest measured corrosion rate. Yttrium, incorporated in the corrosion products, played a role in reducing the corrosion rate across all cases.

The ignition properties of this material exceed 870 °C, therefore, are sufficient even though the ignition temperature was slightly reduced for rapidly solidified samples due to the inevitable oxidation during processing.

**Acknowledgment**

This research was supported by Japan Society for the Promotion of Science (KAKENHI Grant-in-Aid for Scientific Research; 18H05475, 18H05476 and JP20H00312) and MRC International Collaborative Research Grant. The authors would like to thank the Czech Science Foundation (Project No. 22-22248S) and specific university research (A1_FCHT_2024_007) for financial support. The authors acknowledge the assistance provided by the Ferroic Multifunctionalities project, supported by the Ministry of Education, Youth, and Sports of the Czech Republic. Project No. CZ.02.01.01/00/22_008/0004591, co-funded by the European Union. CzechNanoLab project LM2023051 funded by MEYS CR is gratefully acknowledged for the financial support of the measurements/sample fabrication at LNSM Research Infrastructure. Drahomir Dvorsky acknowledges the valuable assistance provided by Soya Nishimoto during the experiments.

**Declaration of generative AI and AI-assisted technologies in the writing process**
During the preparation of this work the author(s) used ChatGPT3.5 in order to check grammar and improve readability. After using this tool/service, the author(s) reviewed and edited the content as needed and take(s) full responsibility for the content of the publication.